\documentstyle[12pt]{article}
\begin{document}

\begin{center}
{\bf HEAVY BARYONS IN A RELATIVISTIC QUARK MODEL\\
WITH A NONLOCAL INTERACTION OF LIGHT QUARKS\\}

\vspace*{1cm}
MIKHAIL IVANOV and VALERY LYUBOVITSKIJ\\
{\it Bogoliubov Laboratory of Theoretical Physics, JINR,\\
Dubna Moscow Region, 141980 Russia\\}
\end{center}

\vspace*{1.5cm}
\begin{abstract}
\noindent Semileptonic decays of bottom and charm baryons are considered
within a relativistic three-quark model with the Gaussian shape
for the baryon-three-quark vertex and standard quark propagators.
\end{abstract}

\section{Introduction}
This paper is focused on the actual problem of heavy flavor physics,
exclusive s.l. decays of low-lying bottom and charm baryons. Recently,
activity in this field has obtained a great interest due to the
experiments worked out by the CLEO Collaboration \cite{CLEO} on the
observation of the heavy-to-light s.l. decay $\Lambda_c^+\to\Lambda e^+\nu_e$.
Also the ALEPH~\cite{ALEPH} and OPAL~\cite{OPAL} Collaborations expect in
the near future to observe the exclusive mode $\Lambda_b\to\Lambda_c\ell\nu$.

In ref. \cite{Aniv} a model for QCD bound states composed from
light and heavy quarks was proposed. Actually, this model is
the Lagrangian formulation of the NJL model with separable interaction
\cite{Goldman} but its advantage consists in the possibility of
studying of baryons as the relativistic systems of three quarks.
The framework was developed for light mesons~\cite{Aniv} and
baryons~\cite{Aniv,PSI}, and also for heavy-light hadrons~\cite{Manchester}.

The purpose of present work is to give a description of properties of baryons
containing a single heavy quark within framework proposed in
ref. \cite{Aniv} and developed in ref. \cite{PSI,Manchester}.
Namely, we report the calculation of observables of semileptonic decays of
bottom and charm baryons: Isgur-Wise functions, asymmetry parameters,
decay rates and distributions.

\section{Model}
Our approach \cite{Aniv} is based on the interaction Lagragians describing
the transition of hadrons into constituent quarks and {\it vice versa}:
\begin{eqnarray}\label{strong}
{\cal L}_B^{\rm int}(x)=g_B\bar B(x)\hspace*{-0.2cm}\int \hspace*{-0.2cm}
dy_1...\hspace*{-0.2cm}\int \hspace*{-0.2cm}dy_3
\delta\left(x-\frac{\sum\limits_i m_iy_i}{\sum\limits_i m_i}\right)
F\left(\sum\limits_{i<j}\frac{(y_i-y_j)^2}{18}\right)
J_B(y_1,y_2,y_3)+h.c.\nonumber
\end{eqnarray}
with $J_B(y_1,y_2,y_3)$ being the 3-quark current with quantum numbers
of a baryon $B$:
\begin{eqnarray}\label{current}
J_B(y_1,y_2,y_3)=\Gamma_1 q^{a_1}(y_1)q^{a_2}(y_2)C\Gamma_2 q^{a_3}(y_3)
\varepsilon^{a_1a_2a_3}\nonumber
\end{eqnarray}
Here $\Gamma_{1(2)}$ are the Dirac matrices, $C=\gamma^0\gamma^2$ is
the charge conjugation matrix, and $a_i$ are the color indices.
We assume that the momentum distribution of the constituents inside
a baryon is modeled by an effective relativistic vertex function
which depends on the sum of relative coordinates only
$F\left(\frac{1}{18}\sum\limits_{i<j}(y_i-y_j)^2\right)$ in the configuration
space where $y_i$ (i=1,2,3) are the spatial 4-coordinates of quarks with
masses $m_i$, respectively. They are expressed through the center of mass
coordinate $(x)$ and relative Jacobi coordinates $(\xi_1,\xi_2)$. The shape
of vertex function is chosen to guarantee ultraviolet convergence of matrix
elements. At the same time the vertex function is a phenomenological
description of the long distance QCD interactions between quarks and gluons.
In the case of light baryons we shall work in the limit of isospin
invariance by assuming the masses of $u$ and $d$ quarks are equal each other,
$m_u=m_d=m$. Breaking of the unitary SU(3) symmetry is taken into account
via a difference of strange and nonstrange quark masses $m_s-m\neq 0$.
In the case of heavy-light baryonic currents we suppose that heavy quark
is much larger than light quark $(m_Q\gg m_{q_1},m_{q_2})$,
i.e. a heavy quark is in the c.m. of heavy-light baryon.
Now we discuss the model parameters. First, there are the baryon-quark
coupling constants and the vertex function in the Lagrangian
${\cal L}_B^{\rm int}(x)$.
The coupling constant $g_B$ is calculated from {\it the compositeness
condition} that means that the renormalization
constant of the baryon wave function is equal to zero,
$Z_B=1-g_B^2\Sigma^\prime_B(M_B)=0$, with $\Sigma_B$ being the baryon mass
operator.
The vertex function is an arbitrary function except that it should make the
Feynman diagrams ultraviolet finite, as we have mentioned above.
We choose in this paper a Gaussian vertex function for simplicity.
In Minkowski space we write $F(k^2_1+k^2_2)=\exp[(k^2_1+k^2_2)/\Lambda_B^2]$
where $\Lambda_B$ is the Gaussian range parameter which is
related to the size of a baryon. It was found \cite{PSI} that for nucleons
$(B=N)$ the value $\Lambda_N=1.25$ GeV gives a good description of the
nucleon static characteristics (magnetic moments, charge radii)
and also form factors in space-like region of $Q^2$ transfer up to 1 GeV$^2$.
In this work we will use the value  $\Lambda_{B_q}\equiv\Lambda_N=1.25$ GeV
for light baryons and consider the value $\Lambda_{B_Q}$ for the heavy-light
baryons as an adjustable parameter.
As far as the quark propagators are concerned we shall use the standard form
of light quark propagator with a mass $m_q$
\begin{eqnarray}\label{Slight}
<0|{\rm T}(q(x)\bar q(y))|0>=
\int{d^4k\over (2\pi)^4i}e^{-ik(x-y)}S_q(k), \,\,\,\,\,\,
S_q(k)={1\over m_q-\not\! k}\nonumber
\end{eqnarray}
and the form
\begin{eqnarray}\label{Sheavy}
S(k+v\bar\Lambda_{\{q_1q_2\}})=
\frac{(1+\not\! v)}{2(v\cdot k+\bar\Lambda_{\{q_1q_2\}}+i\epsilon)}
\nonumber
\end{eqnarray}
for heavy quark propagator obtained in the heavy quark limit (HQL)
$m_Q\to\infty$. The notation are the following:
$\bar\Lambda_{\{q_1q_2\}}=M_{\{Qq_1q_2\}}-m_Q$ is the difference between
masses of heavy baryon $M_{\{Qq_1q_2\}}\equiv M_{B_Q}$
and heavy quark $m_Q$ in the HQL, $v$ is the four-velocity of heavy baryon.
It is seen that the value $\bar\Lambda_{\{q_1q_2\}}$
depends on a flavor of light quarks $q_1$ and $q_2$. Neglecting
the SU(2)-isotopic breaking gives three independent parameters:
$\bar\Lambda\equiv\bar\Lambda_{uu}=\bar\Lambda_{dd}=\bar\Lambda_{du}$,
$\bar\Lambda_{s}\equiv\bar\Lambda_{us}=\bar\Lambda_{ds}$, and
$\bar\Lambda_{ss}$.
Of course, the deficiency of such a choice of light
quark propagator is lack of confinement. This
could be corrected by changing the analytic properties of the propagator.
We leave that to a future study. For the time being we shall
avoid the appearance of unphysical imaginary parts in the Feynman diagrams
by restricting the calculations to the following condition:
the baryon mass must be less than the sum of constituent quark masses
$M_B<\sum\limits_i m_{q_i}$.
In the case of heavy-light baryons the restriction
$M_B<\sum\limits_i m_{q_i}$ trivially gives that the parameter
$\bar\Lambda_{\{q_1q_2\}}$ must be less than the
sum of light quark masses $\bar\Lambda_{\{q_1q_2\}} < m_{q_1}+m_{q_2}$. The
last constraint serves as the upper limit for a choice of parameter
$\bar\Lambda_{\{q_1q_2\}}$.
Parameters $\Lambda_{B_Q}$, $m_s$, $\bar\Lambda$ are fixed in this paper
from the description of data on $\Lambda^+_c\to\Lambda^0+e^+ +\nu_e$ decay.
It is found that  $\Lambda_Q$=2.5 GeV, $m_s$=570 MeV and
$\bar\Lambda$=710 MeV.
Parameters $\bar\Lambda_s$ and $\bar\Lambda_{\{ss\}}$ cannot be adjusted
at this  moment since the experimental data on the decays of heavy-light
baryons having the strange quarks (one or two) are not available. In this
paper we use $\bar\Lambda_s=$850 MeV and $\bar\Lambda_{\{ss\}}=$1000 MeV.

\section{Results}
In this section we give the numerical results for the observables of
semileptonic decays of bottom and charm baryons:
the baryonic Isgur-Wise functions, decay rates and asymmetry parameters.
We check that $\xi_1$ and $\xi_2$ functions are
satisfied to the model-independent Bjorken-Xu inequalities.
Also the description of the $\Lambda^+_c\to\Lambda^0+e^+ +\nu_e$
decay which was recently measured by
CLEO Collaboration \cite{CLEO} is given. In what follows we will use
the following values for CKM matrix elements: $|V_{bc}|$=0.04,
$|V_{cs}|$=0.975.

In our calculations of heavy-to-heavy matrix elements we are restricted
only by one variant of three-quark current for each kind of heavy-light
baryon: {\it Scalar current} for $\Lambda_Q$-type baryons  and
{\it Vector current} for $\Omega_Q$-type baryons \cite{Shuryak,Manchester}.

The functions $\zeta$ and $\xi_1$ have the upper limit
$\Phi_0(\omega)=\frac{\ln(\omega+\sqrt{\omega^2-1})}{\sqrt{\omega^2-1}}$.
It is easy to show that $\zeta(\omega)=\xi_1(\omega)=\Phi_0(\omega)$
when $\bar\lambda=0$. The radii of $\zeta$ and $\xi_1$ have
have the lower bound $\zeta\geq 1/3$ and $\xi_1\geq 1/3$.
Increasing of the $\bar\lambda$ value leads to
the suppression of IW-functions in the physical
kinematical region for variable $\omega$.
The IW-functions $\xi_1$ and $\xi_2$ must satisfy two
model-independent Bjorken-Xu inequalities \cite{Xu}
derived from the Bjorken sum rule for semileptonic $\Omega_b$ decays to
ground and low-lying negative-parity excited charmed baryon states in
the HQL
\begin{eqnarray}
& &1\geq B(\omega)=\frac{2+\omega^2}{3}\xi_1^2(\omega)+
\frac{(\omega^2-1)^2}{3}\xi_2^2(\omega)
+\frac{2}{3}(\omega-\omega^3)\xi_1(\omega)\xi_2(\omega)
\label{ineq1}\\
& &\rho^2_{\xi_1}\geq \frac{1}{3}-\frac{2}{3}\xi_2(1)
\label{ineq2}
\end{eqnarray}
\noindent
The inequality (\ref{ineq2}) for the slope of the $\xi_1$-function
is fulfilled automatically because of $\rho^2_{\xi_1} \geq 1/3$ and
$\xi_2(1) > 0$.
From the inequality (\ref{ineq1})
one finds the upper limit for the function $\xi_1(\omega)$:
$\xi_1(\omega)\leq\sqrt{3/(2+\omega^2)}$

In Fig.1 we plot the $\zeta$ function in the kinematical region
$1\leq \omega \leq \omega_{max}$.
For a comparison the results of other phenomenological
approaches are drawn. There are data of QCD sum
rule~\cite{Grozin}, IMF models~\cite{Kroll,Koerner2},
MIT bag model~\cite{Zalewski}, a simple quark model (SQM)~\cite{Mark1} and
the dipole formula~\cite{Koerner2}. Our result is close to the result of
QCD sum rules~\cite{Grozin}.

In Table 1 our results for total rates are compared with
the predictions of other phenomenological approaches:
constituent quark model \cite{DESY},
spectator quark model \cite{Singleton}, nonrelativistic
quark model \cite{Cheng}.

\newpage

\begin{center}
{\bf Table 1.} Model Results for Rates of Bottom Baryons
(in $10^{10}$ sec$^{-1}$)\\
\end{center}
\begin{center}
\def\arraystretch{1.}
\begin{tabular}{|c|c|c|c|c|} \hline
 Process & Ref. \cite{Singleton} & Ref. \cite{Cheng} & Ref. \cite{DESY}
& Our results\\
\hline\hline
$\Lambda_b^0\to\Lambda_c^+ e^-\bar{\nu}_e$  & 5.9 & 5.1 & 5.14 & 5.39
\\
\hline
$\Xi_b^0\to\Xi_c^+ e^-\bar{\nu}_e$  & 7.2 & 5.3 & 5.21& 5.27
\\
\hline
$\Sigma_b^+\to\Sigma_c^{++} e^-\bar{\nu}_e$
& 4.3 & & & 2.23  \\
\hline
$\Sigma_b^{+}\to\Sigma_c^{\star ++} e^-\bar{\nu}_e$
 & & & &4.56 \\
\hline
$\Omega_b^-\to\Omega_c^0 e^-\bar{\nu}_e$
& 5.4 & 2.3 & 1.52 & 1.87\\
\hline
$\Omega_b^-\to\Omega_c^{\star 0} e^-\bar{\nu}_e$
 & & & 3.41 & 4.01 \\
\hline\hline
\end{tabular}
\end{center}

\vspace*{0.4cm}
Now we consider the heavy-to-light semileptonic modes.
Particular the process $\Lambda^+_c\to\Lambda^0+e^++\nu_e$  which was recently
investigated by CLEO Collaboration \cite{CLEO} is studied in details.
At the HQL ($m_C\to\infty$), the weak hadronic
current of this process is defined by two form factors $f_1$ and $f_2$
\cite{DESY,Cheng}.
Supposing identical dipole forms of the form factors
(as in the model of K\"{o}rner and Kr\"{a}mer \cite{DESY}),
CLEO found that $R=f_2/f_1=$-0.25$\pm$0.14$\pm$0.08. Our form factors have
different $q^2$ dependence. In other words, the quantity $R=f_2/f_1$
has a $q^2$ dependence in our approach. In Fig.10 we plot the results
for $R$ in the kinematical region $1\leq \omega \leq \omega_{max}$ for
different magnitudes of $\bar\Lambda$ parameter.
Here $\omega$ is the scalar product of four velocities of
$\Lambda_c^+$ and $\Lambda^0$ baryons.
It is seen that growth of the $\bar\Lambda$ leads to the
increasing of ratio $R$. The best fit of experimental data is achieved
when our parameters are equal to $m_s=$570 MeV, $\Lambda_Q=$2.5 GeV
and $\bar\Lambda=$710 MeV. In this case the $\omega$-dependence of the
form factors $f_1$, $f_2$ and their ratio $R$ are drawn in Fig.11.
Particularly, we get $f_1(q^2_{max})$=0.8, $f_2(q^2_{max})$=-0.18,
$R$=-0.22 at zero recoil ($\omega$=1 or q$^2$=q$^2_{max}$) and
$f_1(0)$=0.38, $f_2(0)$=-0.06, $R$=-0.16 at maximum recoil
($\omega=\omega_{max}$ or $q^2$=0).
One has to remark that our results at $q^2_{max}$ are closed to the results
of nonrelativistic quark model \cite{Cheng}:
$f_1(q^2_{max})$=0.75, $f_2(q^2_{max})$=-0.17, $R$=-0.23.

Also our result for $R$ weakly deviate from the experimental
data \cite{CLEO} $R=-0.25 \pm 0.14 \pm 0.08$ and the result of
nonrelativistic quark model (Ref. \cite{Cheng}). Our prediction for
the decay rate
$\Gamma(\Lambda^+_c\to\Lambda^0e^+\nu_e)$=7.22$\times$ 10$^{10}$ sec$^{-1}$
and asymmetry parameter $\alpha_{\Lambda_c}$=-0.812 also coincides with the
experimental data $\Gamma_{exp}$=7.0$\pm$ 2.5 $\times$ 10$^{10}$ sec$^{-1}$
and $\alpha_{\Lambda_c}^{exp}$=-0.82$^{+0.09+0.06}_{-0.06-0.03}$ and
the data of Ref. \cite{Cheng} $\Gamma$=7.1 $\times$ 10$^{10}$ sec$^{-1}.$
One has to remark that the success in the reproducing of experimental
results is connected with the using of the $\Lambda^0$ three-quark current
in the $SU(3)$-flavor symmetric form.
By analogy, in the nonrelativistic quark model \cite{Cheng} the assuming
the $SU(3)$ flavor symmetry leads to the presence of the flavor-suppression
factor  $N_{\Lambda_c\Lambda}=1/\sqrt{3}$ in matrix element of
$\Lambda_c^+\to\Lambda^0 e^+\nu_e$ decay. If the $SU(3)$ symmetric
structure of $\Lambda^0$ hyperon is not taken into account the
predicted rate for $\Lambda_c^+\to\Lambda^0 e^+\nu_e$ became too large
(see, discussion in ref. \cite{DESY,Cheng}).
Finally, in Table 2 we give our predictions for some modes of
semileptonic heavy-to-lights transitions.
Also the results of other approaches are tabulated.

\vspace*{0.4cm}
\begin{center}
{\bf Table 2.} Heavy-to-Light Decay Rates (in 10$^{10}$ s$^{-1}$).
\end{center}
\begin{center}
\begin{tabular}{|c|c|c|c|c|c|c|} \hline
 Process & Quantity & Ref.\cite{Singleton} & Ref.\cite{Cheng} &
Ref.\cite{Datta}
& Our & Experiment \\
\hline\hline
$\Lambda_c^+\to\Lambda^0 e^+\nu_e$  & $\Gamma$ & 9.8 & 7.1 &
5.36 & 7.22 & 7.0$\pm$ 2.5 \\
\hline
$\Xi_c^0\to\Xi^- e^+\nu_e$  & $\Gamma$ & 8.5 & 7.4 & & 8.16 & \\
\hline
$\Lambda_b^0\to p e^-\bar\nu_e$ & $\Gamma/|V_{bu}|^2$ & & & 6.48$\times$ 10$^2$ &
7.47$\times$ 10$^2$ &\\
\hline
$\Lambda_c^+\to ne^+\nu_e$  & $\Gamma/|V_{cd}|^2$ & & & &
0.26$\times$ 10$^2$ & \\
\hline\hline
\end{tabular}
\end{center}

\vspace*{.5cm}
\section{Acknowledgements}

We would like to thank J\"{u}rgen K\"{o}rner and Peter Kroll for useful
discussions. This work was supported in part by the INTAS Grant 94-739,
the Heisenberg-Landau Program by the Russian Fund of
Fundamental Research (RFFR)  under contract 96-02-17435-a and the
State Committee of the Russian Federation for
Education (project N 95-0-6.3-67,
Grand Center at S.-Petersburg State University).

\newpage
\listoffigures

\noindent {\bf Fig.1} $\zeta(\omega)$ form factor\\
1. Sutherland \cite{Mark2}\\
2. Grozin \& Yakovlev \cite{Grozin}\\
3. Our result\\
4. Dipole formula\cite{DESY}\\
5. Sadzikowski \& Zalewski \cite{Zalewski}\\
6. K\"{o}rner et al. \cite{DESY}\\
7. Guo \& Kroll \cite{Kroll}\\

\vspace*{1cm}
\noindent {\bf Fig.2} Form factors $f_1, f_2$ and ratio $R$
for $\Lambda^+_c\to\Lambda^0e^+\nu$ decay\\
\end{document}